\documentclass[prd,preprint,a4paper,superscriptaddress,nofootinbib,11pt]{revtex4}
\usepackage{amsmath,amsfonts,amssymb}
\usepackage{txfonts}
\usepackage[T1]{fontenc}
\renewcommand{\d}{\mathrm{d}}
\begin{document}


\begin{center}
{\bf  \Large $\kappa$-Poincar\'{e}-Hopf algebra and  Hopf algebroid 
 structure\\
 of phase space from twist\\}
 
 \bigskip
\bigskip

Tajron Juri\'c  {\footnote{e-mail: tjuric@irb.hr}} \\  
Rudjer Bo\v{s}kovi\'c Institute, Bijeni\v cka  c.54, HR-10002 Zagreb,
Croatia \\[3mm]

Stjepan Meljanac {\footnote{e-mail: meljanac@irb.hr}},
 \\  
Rudjer Bo\v{s}kovi\'c Institute, Bijeni\v cka  c.54, HR-10002 Zagreb,
Croatia \\[3mm] 
 
Rina  \v{S}trajn {\footnote{e-mail: r.strajn@jacobs-university.de}},
\\
Jacobs University Bremen, 28759 Bremen, Germany\\[3mm]

\end{center}
\setcounter{page}{1}


{
We unify $\kappa$-Poincar\'{e} algebra and  $\kappa$-Minkowski spacetime by embedding them into quantum phase space. The
quantum phase space has Hopf algebroid structure to which we apply the twist in order to get $\kappa$-deformed Hopf algebroid structure and $\kappa$-deformed Heisenberg algebra. We explicitly construct $\kappa$-Poincar\'{e}-Hopf algebra and $\kappa$-Minkowski spacetime from twist. It is outlined  how this construction can be extended to $\kappa$-deformed super algebra including exterior derivative and forms. Our results are relevant for constructing physical theories on noncommutative spacetime by twisting Hopf algebroid phase space structure.
}



\newpage

\section{Introduction} 
The noncommutative (NC) spacetime is a  natural setting  for investigating the properties of physical theories  and the structure of spacetime at very small distances \cite{douglas, Szabo}. There are arguments based on quantum gravity \cite{Doplicher, kempf}, and string theory models \cite{Witten, boer}, which suggest that spacetime at Planck length is quantum, i.e. it should be noncommutative. One of the main motivations for dealing with NC spaces is related to the fact that general theory of relativity together with Heisenberg uncertainty principle leads to coordinate uncertainty $\Delta x_{\mu} \Delta x_{\nu} > l^2_{\text{Planck}}$ which can naturally be realized via noncommuting coordinates \cite{Doplicher}. In this setting the spacetime becomes ``fuzzy'' and the notion of smooth spacetime geometry and its symmetry are generalized using Hopf algebraic methods.

$\kappa$-Minkowski spacetime \cite{Lukierski-1}-\cite{rmatrix} is a Lie algebraic deformation of Minkowski spacetime, where $\kappa$ is the deformation parameter usually interpreted as Planck mass or the quantum gravity scale. The full symmetry of special relativity, i.e. Minkowski spacetime is algebraically described by Poincar\'{e}-Hopf algebra $\mathcal{U}(\mathcal{P})$ generated by rotations, boosts and translations. Analogously, the symmetries of $\kappa$-Minkowski spacetime are encoded in the  $\kappa$-Poincar\'{e}-Hopf algebra.  Generalized Poncar\'{e} algebras related to $\kappa$-Minkowski spacetime were considered in \cite{Kovacevic222}.

Some of the main features of physical theories on $\kappa$-Minkowski spacetime are modification of particle statistics \cite{kappaSt}-\cite{Gumesa}, deformed Maxwell's equations \cite{h,hjm11}, Aharonov-Bohm problem \cite{andrade} and quantum gravity effects \cite{dolan, solo, bgmp10, hajume}. The construction of QFT's on NC spaces is of immense importance and is still under investigation \cite{klm00}-\cite{ms11}. $\kappa$-Minkowski spacetime is also related to doubly-special and deformed relativity theories \cite{Kowalski-Glikman-1, Amelino-Camelia-1, Amelino-Camelia-2, Kowalski-Glikman-2, bojowald}.

It is known that the deformations of the symmetry group  can be realized through the application of the Drinfeld twist on that symmetry group \cite{Drinfeld, Drinfeld1}.  The main virtue of the twist formulation is that the deformed (twisted) symmetry algebra is the same as the original undeformed one and the only thing that changes is the coalgebra structure which then leads to the same free field structure as the corresponding commutative field theory.

One of the ideas presented by the group of Wess et al. \cite{a4,a3} is that the symmetries of general relativity, i.e. the diffeomophisms, are considered as the fundamental objects and are deformed using twist \cite{Aschieri,a1}. The generalization of the diffeormorphism symmetry is formulated in the language of Hopf algebras, a setting suitable for studying quantization of Lie groups and algebras. Physical applications of this approach are investigated in \cite{a2}, and especially for black holes in \cite{s2}. Our main motivation is to generalize the ideas of the group of Wess et al. to the notion of the Hopf algebroid \cite{Lu, Bem, inprogress} and to construct both QFT and gravity in Hopf algebroid setting, which is more general and it seams more natural since it deals with the  whole phase space \cite{rmatrix}.

So far there have been attempts in the literature to obtain $\kappa$-Poincar\'{e}-Hopf algebra from Drinfeld twist, but none of them succeeded to accomplish this completely. The Abelian twists \cite{Meljanac-4, ms06, Govindarajan-1}  and Jordanian twits \cite{Borowiec-3}  compatible with $\kappa$-Minkowski spacetime were constructed, but the problem with these twists is that they can not be expressed in terms of the Poincar\'{e} generators and the coalgebra runs out into $\mathcal{U(\mathfrak{igl}(\text{4}))}\otimes\mathcal{U(\mathfrak{igl}(\text{4}))}$.

 In this letter we will show that the key for resolving these problems is to analyze the whole quantum phase space $\mathcal{H}$ and its Hopf algebroid structure. Here, we use the Abelian twist, satisfying cocycle condition. This twist  is not an element of $\kappa$-Poincar\'{e}-Hopf algebra, but an element of  $\mathcal{H}\otimes\mathcal{H}$. By applying the twist to the Hopf algebroid structure of  quantum phase space $\mathcal{H}$ we obtain the Hopf algebroid structure of $\kappa$-deformed Heisenberg algebra $\hat{\mathcal{H}}$. Moreover,  this twist also provides the correct Hopf algebra structure of $\kappa$-Poincar\'{e} algebra when applied to the generators of rotation, boost and momenta.

In section II. we present the $\kappa$-Poincar\'{e}-Hopf algebra in the bicrossproduct basis and show how the generators of $\kappa$-Poincar\'{e}-Hopf algebra and $\kappa$-Minkowski spacetime   can be viewed  as elements of both quantum phase space $\mathcal{H}$ and  $\kappa$-deformed Heisenberg algebra $\hat{\mathcal{H}}$. In section III. we first give the Hopf algebroid structure of quantum phase space $\mathcal{H}$ and then using the twist $\mathcal{F}$  construct the Hopf algebroid structure of $\kappa$-deformed Heisenberg algebra $\hat{\mathcal{H}}$. At the end of section III. we show that the twist $\mathcal{F}$ provides the correct Hopf algebra structure of $\kappa$-Poincar\'{e} algebra. In section IV. we illustrate the construction of exterior derivative and NC one-forms compatible with $\kappa$-Poincar\'{e} algebra and  comment some of the problems which could be resolved  by using the extended twist within super-Heisenberg algebra.

\section{$\kappa$-Poincar\'{e}-Hopf algebra in bicrossproduct basis}

For $\kappa$-Poincar\'{e}-Hopf algebra $\mathcal{U}(\mathcal{P}_{\kappa})$ generated by Lorentz generators $M_{\mu\nu}$ and momentum generators $p_{\mu}$ the coproducts $\Delta$ in bicrossproduct basis \cite{Majid-Ruegg} are\footnote{Greek indices $(\mu,\nu,...)$ are going from 0 to 3, and latin indices $(i,j,...)$ from 1 to 3. Summation over repeated indices is assumed. We use $\eta_{\mu\nu}=\text{diag}(-1,1,1,1).$}
\begin{equation}\begin{split}\label{bicroscoproduct}
 &\Delta p_{0}=p_{0}\otimes 1+1\otimes p_{0},\quad  \Delta p_{i}=p_{i}\otimes 1+e^{a_{0}p_{0}}\otimes p_{i},\\
\Delta M_{i0}=&M_{i0}\otimes 1+e^{a_{0}p_{0}}\otimes M_{i0}-a_{0}p_{j}\otimes M_{ij} , \quad  \Delta M_{ij}=M_{ij}\otimes 1+ 1\otimes M_{ij}\\
\end{split}\end{equation}
where $a_{0}\propto\frac{1}{\kappa}$ is the deformation parameter and $\kappa$ can be interpreted as Planck mass or the quantum gravity scale. 
Equations in (\ref{bicroscoproduct}) describe the coalgebra structure of the $\kappa$-Poincar\'{e}-Hopf algebra and together with antipode $S:\mathcal{U}(\mathcal{P_{\kappa}})\mapsto\mathcal{U}(\mathcal{P_{\kappa}})$ and counit $\epsilon: \mathcal{U}(\mathcal{P_{\kappa}})\mapsto \mathbb{C}$
\begin{equation}\begin{split}
&\epsilon(p_{\mu})=\epsilon(M_{\mu\nu})=0\\
&S(p_{0})=-p_{0} \quad S(p_{i})=-p_{i}e^{-a_{0}p_{0}}\\
&S(M_{ij})=-M_{ij} \quad S(M_{i0})=-e^{-a_{0}p_{0}}\left(M_{i0}+a_{0}M_{ij}p_{j}\right)
\end{split}\end{equation}   make the  $\kappa$-Poincar\'{e}-Hopf algebra. 
The $\kappa$-Poincar\'{e}-Hopf algebra $\mathcal{U}(\mathcal{P}_{\kappa})$ is generated by Lorentz generators $M_{\mu\nu}$ and momentum generators $p_{\mu}$,  where $M_{\mu\nu}$ generate undeformed Lorentz algebra,
\begin{equation}\label{lorentz}
[M_{\mu\nu},M_{\lambda\rho}]=-i\left(\eta_{\nu\lambda}M_{\mu\rho}-\eta_{\mu\lambda}M_{\nu\rho}-\eta_{\nu\rho}M_{\mu\lambda}+\eta_{\mu\rho}M_{\nu\lambda}\right)
\end{equation}
 and $p_{\mu}$ satisfies 
 \begin{equation}\label{pp}
 [p_{\mu},p_{\nu}]=0.
 \end{equation} 
 Additionally, the commutation relations $[M_{\mu\nu},p_{\lambda}]$ are
 \begin{equation}\begin{split}\label{defP}
 &[M_{ij},p_{k}]=i\left(\delta_{ik}p_{j}-\delta_{jk}p_{i}\right), \quad [M_{ij},p_{0}]=0\\
  &[M_{i0},p_{k}]=\delta_{ik}\left(\frac{1-e^{2a_{0}p_{0}}}{2ia_{0}}-\frac{ia_{0}}{2}p^{2}_{l}\right)+ia_{0}p_{i}p_{k}\\
  &[M_{i0},p_{0}]=ip_{i}+ia_{0}p_{i}p_{0}
 \end{split}\end{equation}
Note that in the limit $a_{0}\rightarrow 0$ we have $\Delta\rightarrow\Delta_{0}$, $S\rightarrow S_{0}$ and the  $\kappa$-deformed Poincar\'{e} algebra (\ref{lorentz}-\ref{defP}) reduces to the undeformed Poincar\'{e}-Hopf algebra $\mathcal{U}(\mathcal{P})$.

 In $\kappa$-Minkowski space with deformed coordinates $\{\hat{x}_{\mu}\}$ we have:
\begin{equation}\label{kappa}
[\hat{x}_{i},\hat{x}_{j}]=0, \quad [\hat{x}_{0},\hat{x}_{i}]=ia_{0}\hat{x}_{i}.
\end{equation}
Aiming at covariant notation we define $a_{\mu}=(a_{0},\vec{0})$ and for eq.(\ref{kappa}) get
\begin{equation}\label{kM}
[\hat{x}_{\mu},\hat{x}_{\nu}]=i(a_{\mu}\hat{x}_{\nu}-a_{\nu}\hat{x}_{\mu})
\end{equation}
NC coordinates $\hat{x}$ and momentum $p$ generate the $\kappa$-deformed Heisenberg algebra $\hat{\mathcal{H}}$.

We introduce the action\footnote{For more details see \cite{kovacevic-meljanac}.} 
$\blacktriangleright : \hat{\cal H}(\hat{x},p)\mapsto\hat{\cal A}(\hat{x})$, where $\hat{\cal H}(\hat{x},p)$ is the unital algebra generated by $\hat{x}_{\mu}$ and $p_{\mu}$ and $\hat{\cal A}(\hat{x})$ is a unital subalgebra of $\hat{\cal H}(\hat{x},p)$ generated by $\hat{x}_{\mu}$. This action is defined by the following requirements  
\begin{equation}\begin{split}\label{crnodjelovanje}
&\hat{x}_{\mu} \blacktriangleright \hat{g}(\hat{x})=\hat{x}_{\mu}\hat{g}(\hat{x}),\quad p_{\mu}\blacktriangleright 1=0, \quad M_{\mu\nu}\blacktriangleright 1=0 \\
&p_{\mu}\blacktriangleright \hat{x}_{\nu}=-i\eta_{\mu\nu}, \quad  M_{\mu\nu}\blacktriangleright \hat{x}_{\lambda}=-i\left(\eta_{\nu\lambda}\hat{x}_{\mu}-\eta_{\mu\lambda}\hat{x}_{\nu}\right).\\
\end{split}\end{equation}
Namely, using coproducts (\ref{bicroscoproduct}), action (\ref{crnodjelovanje}) and $G\hat{x}_{\mu}=m(\Delta G(\blacktriangleright\otimes 1)(\hat{x}_{\mu}\otimes 1)), \ \ \forall G\in\mathcal{U}(\mathcal{P}_{\kappa})$\footnote{where $m$ is a multiplication map $m(A\otimes B)=AB$}, one can extract the commutation relations  $[M_{\mu\nu},\hat{x}_{\lambda}]$ and  $[p_{\mu},\hat{x}_{\nu}]$:  
\begin{equation}\label{impuls-xkapa}
[p_{0},\hat{x}_{\mu}]=-i\eta_{0\mu}, \quad [p_{k},\hat{x}_{\mu}]=-i\eta_{k\mu}+ia_{\mu}p_{k},
\end{equation}
\begin{equation}\label{MxLie}
[M_{\mu\nu},\hat{x}_{\lambda}]=-i\left(\eta_{\nu\lambda}\hat{x}_{\mu}-\eta_{\mu\lambda}\hat{x}_{\nu}+a_{\mu}M_{\nu\lambda}-a_{\nu}M_{\mu\lambda}\right).
\end{equation}

The Lorentz generators $M_{\mu\nu}$ can be expressed in terms of $\hat{x}_{\mu}$ and $p_{\mu}$, that is as an element in $\hat{\mathcal{H}}$: 
\begin{equation}\begin{split}\label{raliz}
M_{i0}=&\hat{x}_{i}\left(\frac{e^{2a_{0}p_{0}}-1}{2a_{0}}-\frac{a_{0}}{2}p^{2}_{k}\right)-\hat{x}_{0}p_{i}\\
&M_{ij}=\hat{x}_{i}p_{j}-\hat{x}_{j}p_{i}
\end{split}\end{equation} 
Eqs.\eqref{raliz} are completely compatible with eqs.(1-10).
In this way we have embedded the $\kappa$-Poincar\'{e} algebra (\ref{lorentz}-\ref{defP}) and $\kappa$-Minkowski space (\ref{kappa}) into $\kappa$-deformed Heisenberg algebra $\hat{\mathcal{H}}$. 
 
 The realization for $M_{\mu\nu}$ and $\hat{x}_{\mu}$  in terms of $x_{\mu}$ and $p_{\mu}$, that is as an element in quantum phase space $\mathcal{H}$ (see (\ref{heisenberg})),  corresponding to bicrossproduct basis is :
 \begin{equation}\begin{split}\label{bicrosrealization}
 \hat{x}_{i}&=x_{i}, \quad \hat{x}_{0}=x_{0}-a_{0}x_{k}p_{k}, \\
M_{i0}=x_{i}\left(\frac{Z^2-1}{2a_{0}}-\frac{a_{0}}{2}p^{2}_{k}\right)&-\big(x_{0}-a_{0}x_{k}p_{k}\big)p_{i},\quad M_{ij}=x_{i}p_{j}-x_{j}p_{i},\\
\end{split}\end{equation}
where $Z=e^A$ and  $A=a_{0}p_{0}$  (for more details see \cite{ms06} and \cite{Meljanac-2}). In this way  we have  embedded the $\kappa$-Poincar\'{e} algebra (\ref{lorentz}-\ref{defP})   and $\kappa$-Minkowski space (\ref{kappa})   into quantum phase space $\mathcal{H}$ (which is defined in the next section, see \eqref{heisenberg}).

\section{Quantum phase space and twist}

\subsection{Quantum phase space and  Hopf algebroid structure}
The quantum phase space, i.e. Heisenberg algebra $\mathcal{H}$ is defined by 
\begin{equation}\begin{split}\label{heisenberg}
[x_{\mu}&,x_{\nu}]=[p_{\mu},p_{\nu}]=0\\
&[p_{\mu},x_{\nu}]=-i\eta_{\mu\nu}
\end{split}\end{equation}
 We define the action 
$\triangleright : {\cal H}(x,p)\mapsto {\cal A}(x) $, where $\mathcal{A}$ is a subalgebra, ${\cal A}(x)\subset{\cal H}(x,p)$, generated by $x_{\mu}$. Heisenberg algebra ${\cal H}(x,p)$ can be written as ${\cal H}={\cal A}\;{\cal T}$, where $\mathcal{T}$ is a subalgebra, ${\cal T}(p)\subset{\cal H}(x,p)$, generated by $p_{\mu}$. For any element $f(x)\in{\cal A}(x)$ we have
\begin{equation}\label{djelovanje}
x_{\mu} \triangleright f(x)=x_{\mu}f(x),\quad p_{\mu}\triangleright f(x)=-i\frac{\partial f}{\partial x^{\mu}},
\end{equation}
It is known that for Heisenberg algebra $\mathcal{H}$ there is no Hopf algebra structure. However there exists Hopf algebroid structure defined by undeformed coproduct $\Delta^{\prime}_{0}$ , counit $\epsilon^{\prime}_{0}$ and antipode $S^{\prime}_{0}$
\begin{equation}\begin{split}\label{del0}
\Delta^{\prime}_{0}p_{\mu}=&p_{\mu}\otimes 1+1\otimes p_{\mu}, \quad \Delta^{\prime}_{0}x_{\mu}=x_{\mu}\otimes 1,\\
&\epsilon^{\prime}_{0}(h)=h\triangleright 1,\quad \forall h\in\mathcal{H},\\
S^{\prime}_{0}&(p_{\mu})=-p_{\mu}, \quad  S^{\prime}_{0}(x_{\mu})=x_{\mu}
\end{split}\end{equation}
where we have generated an equivalence class in $\mathcal{A}\otimes \mathcal{A}$ by the relations $\left(\mathcal{R}_{0}\right)_{\mu}\equiv x_{\mu}\otimes 1 - 1\otimes x_{\mu}$ (for details see \cite{rmatrix}). The  $S^{\prime}_{0}$ is antimultiplicative map, $S^{\prime}_{0}: \mathcal{H}\mapsto\mathcal{H}$  and  a generalization of the antipode  $S_{0}$ in Hopf algebra (see \cite{Lu} and \cite{Bem}). The $\epsilon^{\prime}_{0}: \mathcal{H}\mapsto\mathcal{A}$ is a generalization of a  counit $\epsilon_{0}$ in Hopf algebra. If the coproduct $\Delta^{\prime}_{0}$, counit $\epsilon^{\prime}_{0}$ and antipode $S^{\prime}_{0}$ are applied only to the generators of the Poincar\'{e} algebra, they act in a same way as $\Delta_{0}$, $\epsilon_{0}$ and $S_{0}$ and they satisfy the axioms of a Hopf algebra. However, note that: $S^{\prime}_{0}(x_{\mu}p_{\nu})=-p_{\nu}x_{\mu}$ and in $\mathfrak{igl}(4)$-Hopf algebra $S_{0}(x_{\mu}p_{\nu})=-x_{\mu}p_{\nu}$  so for $\mu=\nu$ we have that $S^{\prime}_{0}(x_{\mu}p_{\nu})\neq S_{0}(x_{\mu}p_{\nu})$. The full treatment of Hopf algebroid structure of quantum phase space and $\kappa$-deformed phase space will be presented elsewhere.

\subsection{$\kappa$-deformed phase space from twist}
Let us introduce the twist operator $\mathcal{F}$ (see eq.(61) in \cite{Meljanac-4} and also see \cite{ms06})
\begin{equation}\label{twist}
\mathcal{F}=\text{exp}\left(-iA\otimes x_{k}p_{k}\right)
\end{equation}
where $A=a_{0}p_{0}$, which is Abelian and satisfies the cocycle condition \cite{Govindarajan-1}. We apply it to the Hopf algebroid structure of  quantum phase space $\mathcal{H}$
\begin{equation}\begin{split}
&\Delta^{\prime} h=\mathcal{F}\Delta^{\prime}_{0}h \mathcal{F}^{-1} \\
&\epsilon^{\prime}(h)=m\left\{\mathcal{F}^{-1}(\triangleright\otimes 1)(\epsilon^{\prime}_{0}(h)\otimes 1)\right\}
=h\blacktriangleright 1\\
&S^{\prime}(h)=\chi \ S^{\prime}_{0}(h)\ \chi^{-1} \quad \forall h\in \mathcal{H}
\end{split}\end{equation}
where $\chi^{-1}=m\left[\left(S^{\prime}_{0}\otimes 1\right)\mathcal{F}^{-1}\right]=e^{-iAx_{k}p_{k}}$ and 
$\chi = e^{iAx_{k}p_{k}}$ (note that in the Hopf algebroid $\chi\neq m\left[\left(1 \otimes S^{\prime}_{0}\right)\mathcal{F}\right]$ ).
The results for the generators $x_{\mu}$ and $p_{\mu}$ are
\begin{equation}\begin{split}\label{hopfalg}
\Delta^{\prime} p_{0}=p_{0}\otimes 1+1\otimes p_{0}\equiv\Delta p_{0}& \quad \Delta^{\prime} p_{i}=p_{i}\otimes 1 +e^A\otimes p_{i}\equiv\Delta p_{i}\\
\Delta^{\prime} x_{0}=x_{0}\otimes 1 + 1\otimes a_{0} x_{k}p_{k}& \quad \Delta^{\prime} x_{i}=x_{i}\otimes 1\\
\epsilon^{\prime}(x_{\mu})=\hat{x}_{\mu}& \quad \epsilon^{\prime}(p_{\mu})=0\\
S^{\prime}(p_{0})=-p_{0}\equiv S(p_{0})& \quad S^{\prime}(p_{i})=-p_{i}e^{-A}\equiv S(p_{i})\\
S^{\prime}(x_{0})=x_{0}-a_{0}x_{k}p_{k}& \quad  S^{\prime}(x_{i})=x_{i}e^{A}
\end{split}\end{equation}
 where we have generated an equivalence class in $\left(\mathcal{A}\otimes \mathcal{A}\right)_{\mathcal{F}}$ by the relations $\mathcal{R}_{\mu}=\mathcal{F}(\mathcal{R}_{0})_{\mu}\mathcal{F}^{-1}$ (for details see \cite{rmatrix}). We point out that  the coproduct $\Delta^{\prime}$ and antipode $S^{\prime}$ reduce to $\Delta$ and $S$ when applied to the generators of $\kappa$-Poincar\'{e} algebra (see subsection C). 
With the coproduct $\Delta^{\prime}$, antipode $S^{\prime}$ and counit $\epsilon^{\prime}$ given in (\ref{hopfalg}), we have defined the $\kappa$-deformed Hopf algebroid structure of quantum phase space $\mathcal{H}$.

The NC coordinates $\hat{x}$ are given by
\begin{equation}\begin{split}
&\hat{x}_{i}=m\left(\mathcal{F}^{-1}(\triangleright\otimes 1)(x_{i}\otimes 1)\right)=x_{i}\\
&\hat{x}_{0}=m\left(\mathcal{F}^{-1}(\triangleright\otimes 1)(x_{0}\otimes 1)\right)=x_{0}-a_{0}x_{k}p_{k}
\end{split}\end{equation} 
and satisfy  the $\kappa$-Minkowski algebra (\ref{kappa}). Hence, from equations  (\ref{hopfalg}) we have 
\begin{equation}\begin{split}
&\Delta^{\prime} \hat{x}_{\mu}=\hat{x}_{\mu}\otimes 1 \quad  \hat{\epsilon}^{\prime}(\hat{x}_{\mu})=\hat{x}_{\mu}\blacktriangleright 1=\hat{x}_{\mu}\\
&S^{\prime}(\hat{x}_{i})=\hat{x}_{i}e^{A} \quad S^{\prime}(\hat{x}_{0})= \hat{x}_{0}+a_{0}p_{k}\hat{x}_{k}
\end{split}\end{equation}
which  defines the Hopf algebroid structure of $\kappa$-deformed Heisenberg algebra $\hat{\mathcal{H}}$.

\subsection{$\kappa$-Poincar\'{e}-Hopf algebra from twist}
The realization of Lorentz generators $M_{\mu\nu}$ is given in terms of $\mathcal{H}$ in (\ref{bicrosrealization}). Then using the eq.(\ref{del0}) for coproduct and homomorphism of $\Delta^{\prime}_{0}$ we can calculate $\Delta^{\prime}_{0}M_{i0}=\Delta^{\prime}_{0}x_{i}\Delta^{\prime}_{0}\left(\frac{Z^2-1}{2a_{0}}-\frac{a_{0}}{2}p^{2}_{k}\right)-\Delta^{\prime}_{0}\hat{x}_{0}\Delta^{\prime}_{0}p_{i}$ and $\Delta^{\prime}_{0}M_{ij}=\Delta^{\prime}_{0}x_{i}\Delta^{\prime}_{0}p_{j}-\Delta^{\prime}_{0}x_{j}\Delta^{\prime}_{0}p_{i}=M_{ij}\otimes 1 + 1\otimes M_{ij}\equiv \Delta_{0}M_{ij}$. Note that $\Delta^{\prime}_{0}M_{i0}\neq M_{i0}\otimes 1 + 1\otimes M_{i0} $. Furthermore, applying the twist $\mathcal{F}$ we get the deformed coproduct
\begin{equation}\begin{split}
\Delta^{\prime} M_{ij}&=\mathcal{F}\Delta^{\prime}_{0}M_{ij}\mathcal{F}^{-1}=M_{ij}\otimes 1+1\otimes M_{ij}\equiv\Delta M_{ij}\\
\Delta^{\prime} M_{i0}&=\mathcal{F}\Delta^{\prime}_{0}M_{i0}\mathcal{F}^{-1}=\Delta^{\prime} x_{i}\Delta^{\prime}\left(\frac{Z^2-1}{2a_{0}}-\frac{a_{0}}{2}p^{2}_{k}\right)-\Delta^{\prime}\hat{x}_{0}\Delta^{\prime} p_{i}\\
=&(x_{i}\otimes 1)\left(\frac{Z^2 \otimes Z^2 -1\otimes 1}{2a_{0}}-\frac{a_{0}}{2}\Delta^{\prime}(p^{2}_{k})\right)-(\hat{x}_{0}\otimes 1)\Delta^{\prime}p_{i}\\
&=M_{i0}\otimes 1+e^{a_{0}p_{0}}\otimes M_{i0}-a_{0}p_{j}\otimes M_{ij}\equiv\Delta M_{i0}
\end{split}\end{equation}
where in the last line we have used $x_{i}\otimes 1=Z^{-1}\otimes x_{i}$ which is derived from $\mathcal{R}_{i}$, \cite{rmatrix}.
Similarly, using (\ref{hopfalg}) and antihomomorphism of the antipode $S^{\prime}_{0}$ we get
\begin{equation}\begin{split}
S^{\prime}(M_{ij})&=\chi\ S^{\prime}_{0}(M_{ij})\ \chi^{-1}=S^{\prime}(p_{j})S^{\prime}(x_{j})-S^{\prime}(p_{i})S^{\prime}(x_{j})=-M_{ij}\equiv S(M_{ij})\\
S^{\prime}(M_{i0})&=\chi\ S^{\prime}_{0}(M_{i0})\ \chi^{-1}=S^{\prime}\left(\frac{Z^2-1}{2a_{0}}-\frac{a_{0}}{2}p^{2}_{k}\right)S^{\prime}(x_{i})-S^{\prime}(p_{i})S^{\prime}(\hat{x}_{0})\\
&=-e^{-a_{0}p_{0}}\left(M_{i0}+a_{0}M_{ij}p_{j}\right)\equiv S(M_{i0})
\end{split}\end{equation}
Hence, the twist $\mathcal{F}$ in (\ref{twist}) applied to the Hopf algebroid structure gives the correct Hopf algebra structure of $\kappa$-Poincar\'{e} algebra generated by $M_{\mu\nu}$ and $p_{\mu}$, eqs. (\ref{bicroscoproduct}-\ref{defP}). 

In this letter we have presented the $\kappa$-Poincar\'{e}-Hopf algebra in the  bicrossproduct basis from twist (\ref{twist}). This can be generalized to the twist corresponding to any realization \cite{inprogress}. The flip operator and universal $R$-matrix were analyzed in \cite{rmatrix}.
    
\section{Exterior derivative, NC forms and $\kappa$-deformed super algebra}
We define the exterior derivative $\hat{\text{d}}$ and the NC one form $\hat{\xi}_{\mu}$ in a  usual way \cite{Sitarz, Meljanac-1, Meljanac-2, MKJj, diff}   
\begin{equation}\label{forms}
\hat{\text{d}}^2=0, \quad \hat{\xi}_{\mu}=[\hat{\text{d}},\hat{x}_{\mu}]
\end{equation}
with the properties 
\begin{equation}
\left\{\hat{\xi}_{\mu},\hat{\xi}_{\nu}\right\}=0, \quad \left\{\hat{\d}, \hat{\xi}_{\mu}\right\}=0
\end{equation}
and the commutator between NC forms and NC coordinates $[\hat{\xi}_{\mu},\hat{x}_{\nu}]$ has to be constructed in order to be consistent with (super)Jacobi identities and in doing so, the most general expression reads 
\begin{equation}
[\hat{\xi}_{\mu},\hat{x}_{\nu}]=K_{\mu\nu}^{\lambda}(p)\hat{\xi}_{\lambda}
\end{equation}
The equation for $\kappa$-Minkowski space (\ref{kM}) together with (\ref{forms}) implies the consistency condition
\begin{equation}\label{condition}
[\hat{\xi}_{\mu}, \hat{x}_{\nu}] -[\hat{\xi}_{\nu}, \hat{x}_{\mu}]=i(a_{\mu}\hat{\xi}_{\nu}-a_{\nu}\hat{\xi}_{\mu})
\end{equation}
In order to have a differential calculus compatible with $\kappa$-Poincar\'{e} algebra we  require
\begin{equation}\begin{split}
&[M_{\mu\nu},\hat{\text{d}}]=[p_{\mu},\hat{\text{d}}]=[p_{\mu},\hat{\xi}_{\nu}]=0\\
&[M_{\mu\nu},\hat{\xi}_{\lambda}]=-i\left(\eta_{\nu\lambda}\hat{\xi}_{\mu}-\eta_{\mu\lambda}\hat{\xi}_{\nu}\right)\\
\end{split}\end{equation}
Hence, we have unified the algebra generated by  $\left\{\hat{x},p,M_{\mu\nu},\hat{\xi}, \hat{\d}\right\}$ in a super-algebra.

Sitarz in \cite{Sitarz} constructed differential algebra $\left\{\hat{\xi},\hat{x},\phi\right\}$ which can be exactly realized in the bicrossproduct basis (\ref{bicrosrealization}) as:
\begin{equation}\begin{split}\label{realizatSitarz}
\hat{\xi}_{0}=&\xi_{0}\Big(2-\frac{a^{2}_{0}}{2}p^{2}_{i}Z^{-1}-\text{cosh}(A)\Big)-a_{0}\xi_{k}p_{k}\\
&\ \ \ \hat{\xi}_{k}=\xi_{k}+a_{0}\xi_{0}p_{k}Z^{-1},\\
\phi =&-\hat{\text{d}}_{s}=\xi_{0}\Big(\frac{i}{a_{0}}\text{sinh}(A)-\frac{ia_{0}}{2}p^{2}_{i}Z^{-1}\Big) \\
\end{split}\end{equation}
However, we point out that there does not exist realization for the generators $M_{\mu\nu}$  compatible with $[M_{\mu\nu},\hat{\text{d}}]=0$ and Lorentz algebra (\ref{lorentz}).

In the papers \cite{MKJj,diff} it is shown that the above super algebra has no realization if $\hat{x}_{i}=x_{i}$ and $\hat{x}_{0}=x_{0}-a_{0}x_{k}p_{k}$.

The twist $\mathcal{F}$ defined in (\ref{twist}) leads to $\hat{\xi}_{\mu}=m\left(\mathcal{F}^{-1}(\triangleright\otimes 1)( \xi_{\mu}\otimes 1\right)=\xi_{\mu}$ and this is in contradiction with the consistency condition (\ref{condition}). The analysis of these problems is presented in \cite{diff}, using super Heisenberg algebra and the extended twist $\mathcal{F}_{ext}$. There we have constructed a bicovariant calculus compatible with $\kappa$-deformed $\mathfrak{igl}$(4)-Hopf algebra, for which the consistency condition is satisfied. The physical consequences of this construction is still a work in progress and will be presented elsewhere.\\

In this letter we have analyzed the whole phase space and its Hopf algebroid structure. The idea is to construct quantum field theory and gravity in the Hopf algebroid setting generalizing the ideas
presented by the group of Wess et al. In \cite{fluid, fluid1} NC fluid was analyzed.  Using the realization formalism for Snyder space, fluid equations of motion and their perturbative solutions were derived. It is of interest to generalize this approach using realization of $\kappa$-deformed phase space.  This letter should be considered as a starting point for a further investigation along this lines (work in progress).

\bigskip

\noindent{\bf Acknowledgment}\\
We would like to thank Domagoj Kova\v{c}evi\'{c}, Anna Pachol, Andjelo Samsarov,  and Zoran \v{S}koda for useful discussions.
This work was supported by the Ministry of Science and Technology of the Republic
of Croatia under contract No. 098-0000000-2865. R.\v{S}. gratefully acknowledges support from the DFG within the Research Training Group 1620 ``Models of Gravity''.\\

\end{document}